\documentclass[
prd
,showpacs,amssymb,superscriptaddress,aps
]{revtex4}
\usepackage{graphicx}
\usepackage{color}
\input{epsf}

\usepackage{amsmath,amssymb}
\usepackage{bm}
\usepackage{times}
\usepackage{ulem}
\usepackage{hyperref}
\hypersetup{
    colorlinks=true,
    linkcolor=blue,
    filecolor=magenta,      
    urlcolor=cyan,
    citecolor=blue
}
\newcommand{\dalm}{\kern1pt\vbox{\hrule height 0.9pt\hbox{\vrule width 0.9pt
\hskip 2.5pt\vbox{\vskip 5.5pt}\hskip 3pt\vrule width 0.3pt}\hrule height 0.3pt}
\kern1pt}

\newcommand{\lsim}{\, \, \raisebox{-0.8ex}{$\stackrel{\textstyle <}{\sim}$ }}

\begin{document}

%\twocolumn[\hsize\textwidth\columnwidth\hsize\csname @twocolumnfals\endcsname

% For two column
%\wideabs{

\title{Accuracy of relativistic Cowling approximation in protoneutron star asteroseismology}

\author{Hajime Sotani}
\email{sotani@yukawa.kyoto-u.ac.jp}
\affiliation{Astrophysical Big Bang Laboratory, RIKEN, Saitama 351-0198, Japan}
\affiliation{Interdisciplinary Theoretical \& Mathematical Science Program (iTHEMS), RIKEN, Saitama 351-0198, Japan}

\author{Tomoya Takiwaki}
\affiliation{Division of Science, National Astronomical Observatory of Japan, 2-21-1 Osawa, Mitaka, Tokyo 181-8588, Japan}
\affiliation{Center for Computational Astrophysics, National Astronomical Observatory of Japan, 2-21-1 Osawa, Mitaka, Tokyo 181-8588, Japan}

\date{\today}

% Abstract
\begin{abstract}
The relativistic Cowling approximation, where the metric perturbations are neglected during the fluid oscillations, is often adopted for considering the gravitational waves from the protoneutron stars (PNSs) provided via core-collapse supernova explosions. In this study, we evaluate how the Cowling approximation works well by comparing the frequencies with the Cowling approximation to those without the approximation. Then, we find that the behavior of the frequencies with the approximation is qualitatively the same way as that without the approximation, where the frequencies with the approximation can totally be determined within $\sim 20\%$ accuracy. In particular, the fundamental mode with the Cowling approximation is overestimated. 
In addition, we also  discuss the damping time of various eigenmodes in gravitational waves from the PNSs, where the damping time for the PNSs before the avoided crossing between the $f$- and $g_1$-modes, is quite different from that for cold neutron stars, but it is more or less similar to that for cold neutron stars in the later phase. The damping time is long enough compared to the typical time interval of short-Fourier transformation that often used in the analysis, and that ideally guarantees the validity of the transformation.
\end{abstract}

%
%%%%%%%%%%%%%%%%%%%%%%%%%%%%%%%%%%%%%%%%%%%%%%%%%
%  04.30.-w  :  Gravitational waves
%  04.40.Dg :  Relativistic stars: structure, stability, and oscillations (see also 97.60.-s Late stages of stellar evolution) 
%  21.65.Ef  :  Symmetry energy
%  26.60.Gj  :  Neutron star crust
%  21.60.-n  :   Nuclear structure models and methods
%  97.10.Sj  :   Pulsations, oscillations, and stellar seismology 
%%%%%%%%%%%%%%%%%%%%%%%%%%%%%%%%%%%%%%%%%%%%%%%%%
%]
% For two column
%}
\maketitle

%\baselineskip 24pt

%%%%%%%%%%%%%%%%%%%%%%%%%%%%%%%%%%%%%%%%%%%%%%%%
\section{Introduction}
\label{sec:I}
%%%%%%%%%%%%%%%%%%%%%%%%%%%%%%%%%%%%%%%%%%%%%%%%

Gravitational waves detected from the compact binary mergers have opened a multi-messenger era, where the gravitational waves become a new tool to see the universe together with the electromagnetic waves and neutrino signals. In fact, at the event of GW170817, which is considered as a binary neutron star merger, not only the gravitational waves but also the electromagnetic counterparts have been detected \cite{GW6,EM}. In addition to the current gravitational wave detectors, such as the advanced LIGO (Large Interferometer Gravitational-wave Observatory) and advanced Virgo, the Japanese detector, i.e., KAGRA \cite{aso13} will join the network of  gravitational wave detectors soon. Moreover, the third-generation gravitational wave detectors, such as the Einstein Telescope and Cosmic Explorer \cite{punturo,CE}, have also been discussed as the next step. These attempts may enable us to detect the gravitational waves from the core-collapse supernova, which is the last moment of a massive star, together with the compact binary mergers.

The gravitational waves from the supernova explosions have mainly been studied via numerical simulations  performed under different assumptions about symmetries with various progenitor models and equations of state (EOSs) (e.g., \cite{Murphy09,MJM2013,Ott13,CDAF2013,Yakunin15,KKT2016,Andresen16,Richers2017,Takiwaki2017,OC2018,RMBVN19,VBR2019,PM2020}). Through these studies, an understanding of a characteristic supernova gravitational wave signal in numerical simulations is beginning to emerge, where the frequency increases with time after core-bounce from several hundred hertz up to around a few kilohertz. This ramp up signal is initially considered to correspond to the Brunt-V\"{a}is\"{a}l\"{a} frequency at the protoneutron star (PNS) surface \cite{MJM2013,CDAF2013}, where it is sometimes called the surface gravity ($g$-) mode. Nevertheless, since the Brunt-V\"{a}is\"{a}l\"{a} frequency is  locally determined with the PNS properties, which strongly depend on the definition of the PNS surface, and generally not associated with the global oscillation frequency of the system, the ramp up gravitational wave signal shown in the numerical simulations should come from a kind of global oscillations of the system, such as the core region of the progenitor or the PNS itself. In fact, it is shown that the ramp up signals qualitatively correspond to the fundamental ($f$-) mode oscillation of the PNS \cite{MRBV2018,SKTK2019,ST2020b} or the $g_2$-mode oscillation (with different classification) of the region between the stellar center and the shock radius \cite{TCPF2018,TCPOF2019a,TCPOF2019b}. Meanwhile, recently it is reported that the dominant source of the gravitational waves comes from Ledoux convection rather than the $f$- or $g$-modes, investigating a three-dimensional simulation with neutrino and hydrodynamics code {\small CHIMERA} \cite{Mezzacappa20}. Definitely,  further study is necessary to identify the origin.

Anyway, in order to find the correspondence between the gravitational wave signals shown in the numerical simulations and the PNS properties, the perturbative approach, the so-called asteroseismology, is also one of the crucial ways especially for the case of core-collapse supernovae, because the supernova explosions are more or less similar to spherically symmetric objects, compared to the compact binary mergers, which lead to the less energetic gravitational wave radiations. Asteroseismology is a similar way to seismology for the Earth and helioseismology for the Sun, where one can extract the properties of the objects via their oscillation frequency as an inverse problem. For compact objects, not only the electromagnetic waves but also the gravitational waves become important information for adopting asteroseismology \cite{KS1999}. So far, asteroseismology for cold neutron stars has been studied well, where one examines the specific frequencies on the background neutron star models. For instance, it is suggested that the crust properties of neutron stars are constrained by identifying the quasi-periodic oscillations observed in the giant flares with the crustal torsional oscillations (e.g., \cite{GNHL2011,SNIO2012,SIO2016}). In addition, the direct observations of the gravitational waves from compact objects may tell us their mass, radius, and EOS (e.g., \cite{AK1996,AK1998,STM2001,SH2003,SYMT2011,PA2012,DGKK2013,Sotani2020}).

Compared to the extensive studies for cold neutron stars, the PNS asteroseismology is relatively poor. This may partially come from the difficulty for preparing the background PNS models, on which the linear analysis is done. That is, in order to construct the PNS models, one has to know the radial distributions of the entropy and electron fraction together with the pressure and energy density, but these properties can be determined via the numerical simulations of the core-collapse supernova with an appropriate progenitor model. Owing to the development of the numerical simulations, the number of the studies about PNS asteroseismology gradually increases \cite{MRBV2018,SKTK2019,ST2020b,TCPF2018,TCPOF2019a,TCPOF2019b,FMP2003,Burgio2011,FKAO2015,ST2016,Camelio17,SKTK2017,SS2019,WS2019,ST2020a}.

These studies about the PNS asteroseismology are broadly separated into two groups by the difference of the  background models adopted for the linear analysis. One is that the PNS surface is determined at the specific density \cite{MRBV2018,SKTK2019,ST2020b,FMP2003,Burgio2011,FKAO2015,ST2016,Camelio17,SKTK2017,SS2019,ST2020a}, where the boundary condition imposed at the PNS surface is the same as in the standard asteroseismology, i.e., the Lagrangian perturbation of pressure should be zero. For the PNS asteroseismology, since the matter density beyond the surface of the PNS is nonzero, contrary to the assumptions usually employed in asteroseismology, this boundary condition is totally an assumption, but one can basically classify the eigenmodes according to the standard asteroseismology. As a disadvantage of this approach, the eigenmodes may depend on the surface density, but it is found that at least the $f$- and $g_1$-modes are almost independent on the surface density in the later phase after core-bounce, e.g., after $\sim 500$ ms \cite{MRBV2018,ST2020b}. Another is that the region inside the shock radius is considered as a background for linear analysis \cite{TCPF2018,TCPOF2019a,TCPOF2019b,WS2019}, where the boundary condition imposed at the shock radius is that the radial Lagrangian displacement should be zero. In this case, since the boundary condition is completely different from that for the standard asteroseismology, i.e., the problem to solve is mathematically different, one has to reclassify the eigenmodes. Through these attempts, the ramp up signal shown in the numerical simulations is found to correspond to the $f$-mode oscillation of the PNS \cite{MRBV2018,SKTK2019,ST2020b} or the $g_2$-mode oscillation (with a new classification) of the region inside the shock radius \cite{TCPF2018,TCPOF2019a,TCPOF2019b}. We remark that the $f$- and $g$-modes are mainly discussed in these studies, but the pressure ($p$-) modes, which are also acoustic waves and overtone of the $f$-mode, must play an important role in gravitational wave aseteroseismology if observed \cite{KS1999}.

In most of these studies about PNS asteroseismology, the relativistic Cowling approximation has been adopted for simplicity, where the metric perturbations are neglected during the fluid oscillations, even though the metric perturbations are generally coupled with the fluid perturbations in a relativistic framework. For cold neutron stars, the study for checking the accuracy of the approximation has been done, where it is shown that the frequencies with the approximation are overestimated by less than $\sim 20\%$ for typical neutron stars \cite{YK1997}. On the other hand, for PNSs there are only a few studies, where some terms (not all terms) of the metric perturbations are taken into account, and the resultant frequencies are compared to those with the Cowling approximation \cite{MRBV2018,TCPOF2019a}. In Ref. \cite{MRBV2018} the perturbation of the lapse function is taken into account with the PNS model determined by the specific density, where it is shown that at least the $f$-mode frequency with the approximation is underestimated, while in Ref. \cite{TCPOF2019a} the perturbations of the lapse function and conformal factor are taken into account for the region inside the shock radius, where it is shown that  the $g_2$-mode frequency (corresponding to the ramp up signal) with the approximation is overestimated. So, both results seem not to be consistent. In addition, in Refs. \cite{FMP2003,Burgio2011,Camelio17} all terms of metric perturbations are taken into account for the PNS models whose baryonic mass is fixed during the evolution, i.e., the infalling accretion matter is neglected, but the comparison to the frequencies with Cowling approximation has not been made. So, in this paper, we will calculate the gravitational wave frequencies without the Cowling approximation, where we take into account all terms of metric perturbations, and examine how accurate the approximation works. In addition, we also discuss the damping time of each mode, which is the advantage without the Cowling approximation.

This paper is organized as follows. In Sec. \ref{sec:PNSmodel}, we describe the PNS models considered in this study. In Sec. \ref{sec:Oscillation}, we show the eigenfrequencies of gravitational waves from the PNS without the Cowling approximation, compare them to those with the Cowling approximation, and show the accuracy of the Cowling approximation. In this section, we also discuss the damping time for the eigenmodes of gravitational waves.  Then, we make a conclusion in Sec. \ref{sec:Conclusion}. Unless otherwise mentioned, we adopt geometric units in the following, $c=G=1$, where $c$ denotes the speed of light, and the metric signature is $(-,+,+,+)$.

%%%%%%%%%%%%%%%%%%%%%%%%%%%%%%%%%%%%%%%%%%%%%%%%
\section{PNS Models}
\label{sec:PNSmodel}
%%%%%%%%%%%%%%%%%%%%%%%%%%%%%%%%%%%%%%%%%%%%%%%%

In order to make a linear analysis, first, one has to prepare a background PNS model. In this study, we adopt the same PNS models discussed in our previous study \cite{ST2020b}. That is, we adopt the $2.9M_\odot$ helium star (He2.9) as a progenitor model \cite{He29} together with LS220 \cite{LS220} as the EOS in a high density region. LS220 is constructed with the incompressibility $K_0=220$ MeV and the slope parameter of the nuclear symmetry energy $L=75.8$ MeV, with which the expected maximum mass for a cold neutron star is $2.0M_\odot$. Then, two-dimensional hydrodynamical simulation has been done, adopting {\small 3DnSNe} code, where the neutrino transport is solved by isotropic diffusion source approximation \cite{liebendoerfer2009,takiwaki2014}. We remark that with the same numerical code, one-, two-, and three-dimensional hydrodynamic equations for core-collapse supernovae have also been solved \cite{takiwaki2016,oconnor2018,kotake2018,nakamura2019,sasaki2019,zaizen2019}. In practice, we employ the resolution in the spherical polar grid of $512\times 128$, with which the radial grid covers from the stellar center up to $5000$ km and the polar grid covers in the range between 0 and $\pi$. With respect to the set of neutrino reactions, we adopt the same procedure adopted in Ref. \cite{kotake2018}. Once the two-dimensional numerical simulation has been done, the PNS models are prepared by averaging the physical properties in the angular ($\theta$) direction \cite{note}. The PNS surface is determined at the surface density $\rho_{\rm s}=10^{11}$ g/cm$^3$. In this study, we especially consider the PNS models for the calculation of the gravitational wave frequencies without the Cowling approximation (with full perturbations) at each 0.1 sec after core-bounce, i.e., $T_{\rm pb}=0.1$, 0.2, 0.3, 0.4, 0.5, 0.6, 0.7, 0.8, and 0.9 sec, where $T_{\rm pb}$ denotes the time after core-bounce.

%%%%%%%%%%%%%%%%%%%%%%%%%%%%%%%%%%%%%%%%%%%%%%%%
\section{Gravitational wave frequencies from PNS}
\label{sec:Oscillation}
%%%%%%%%%%%%%%%%%%%%%%%%%%%%%%%%%%%%%%%%%%%%%%%%
Before going to our result, we briefly mention about our method. The oscillations on a spherically symmetric background can be decomposed, using the spherical harmonics, $Y_{\ell k}(\theta,\phi)$, with the azimuthal quantum number $\ell$ and the magnetic quantum number $k$, where the oscillations can be classified into two families by its parity, i.e., axial and polar. In this study, we focus on the  polar-type oscillations with $\ell=2$ and $k=0$, because $\ell=2$ mode is dominant in the gravitational wave radiation, unlike the axial-type oscillations the polar-type oscillations involve the density variation (see Appendix \ref{sec:appendix_1} for the tangible expression of the oscillations with polar parity), and the dependence on $k$ degenerates into $k=0$ due to the nature of spherically symmetric background.

Without the Cowling approximation, i.e., taking into account the metric perturbations together with the fluid perturbations, the perturbation equations can be derived by linearizing the the equations of general relativistic hydrodynamics (Einstein equations and conservation of stress-energy). The concrete equation system, the boundary conditions, and how to determine the gravitational wave frequency are basically the same procedure as in Ref. \cite{STM2001}. That is, we just adopt the PNS models as a background in this study,  instead of cold neutron star models with density discontinuity in Ref. \cite{STM2001}. The perturbation equations inside and outside the PNSs are briefly shown in Appendix \ref{sec:appendix_1} and \ref{sec:appendix_2}, respectively, while the boundary conditions and how to determine the QNMs are shown in Appendix \ref{sec:appendix_3} and \ref{sec:appendix_4}, in which the equations are completely the same as in Ref. \cite{STM2001}. Eventually, the problem to solve becomes an eigenvalue problem with respect to the eigenvalue $\omega$.

\begin{figure}[tbp]
\begin{center}
\includegraphics[scale=0.5]{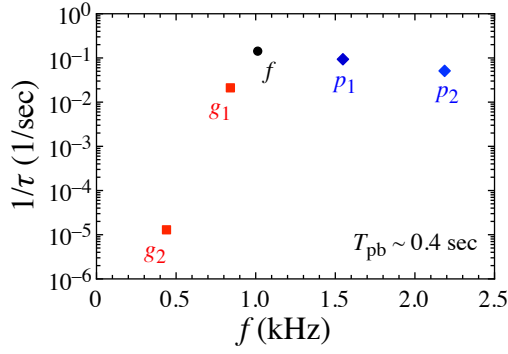} 
\end{center}
\caption{%%
The eigenvalue of the oscillation mode in non-Cowling treatment has complex value. For the PNS model at $T_{\rm pb}\simeq 0.4$ sec, several quasi-normal modes are shown. The real part is the frequency ($f$), while the imaginary part is the damping rate ($1/\tau$).
}%%
\label{fig:ReIm-400}
\end{figure}

The calculated eigenvalue, $\omega$, is a complex value because the gravitational waves would carry out the oscillation energy. So, the eigenmodes are called  quasi-normal modes (QNMs). The real and imaginary parts of $\omega$ are associated with the oscillation frequency, $f$, via ${\rm Re}(\omega)=2\pi f$ and the damping time, $\tau$, via ${\rm Im}(\omega)=1/\tau$, respectively. As an example, for the PNS model at $T_{\rm pb}=0.4$ sec, the frequency and damping rate ($1/\tau$) for several QNMs are shown in Fig. \ref{fig:ReIm-400}, where the circle, diamonds, and squares denote the $f$-, $p_i$-, and $g_i$-modes for $i=1$ and 2, respectively. In general, there are an infinite number of $p_i$- and $g_i$-modes, where the positive frequency of $p_i$-mode ($g_i$-mode) increases (decreases) as $i$ increases. So, the $p_i$-modes for $i\ge 3$ appear on the right side of the $p_2$-mode, while the $g_i$-modes for $i\ge 3$ appear on the left side of the $g_2$-mode (and are more than 0) in this figure, although we do not consider in this study. In this study, we adopt the standard mode classification, i.e., the $f$-mode has no node in the eigenfunction, while the $p_i$- and $g_i$-modes have $i$ nodes in the corresponding eigenfunction. We remark that the QNMs associated with the oscillations of spacetime itself, i.e., the so-called $w$-modes, which are almost independent of the fluid oscillations, exist in the relativistic framework, in addition to the fluid modes considered in this study, such as the $f$-, $p_i$-, and $g_i$-modes. Unlike the fluid modes, the damping rate of $w$-mode is comparable to the oscillation frequency. Anyway, since the $w$-mode damping rate is comparable to its frequency \cite{FMP2003,SKTK2017}, the extraction of the $w$-modes from the gravitational wave signal may be more difficult.

%%%%%%%%%%%%%%%%%%%%%%%%%%%%%%%%%%%
% Figure 1
%%%%%%%%%%%%%%%%%%%%%%%%%%%%%%%%%%%
%%%%%%%%%%%%%%%%%%%%%%%%%%%%%%%%%%%

\begin{figure}[tbp]
\begin{center}
\includegraphics[scale=0.5]{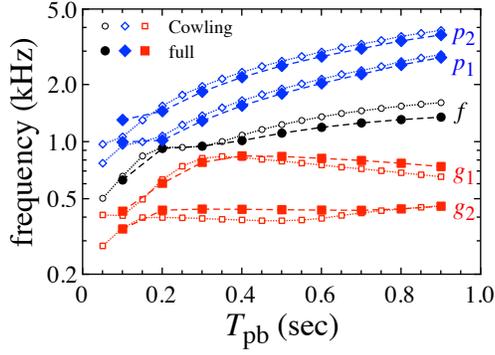} 
\end{center}
\caption{%%
Time evolution of the gravitational wave frequency for the $f$-, $p_1$-, $p_2$-, $g_1$-, and $g_2$-modes. The filled marks with dashed lines correspond to the results obtained without the Cowling approximation (full perturbations) in this study, while the open marks with dotted lines correspond those with the Cowling approximation in Ref. \cite{ST2020b}.
}%%
\label{fig:ft}
\end{figure}

On the other hand, the perturbation equations with the Cowling approximation are derived by linearizing the energy-momentum conservation law. The concrete equations and the boundary conditions imposed at the stellar center and at the PNS surface are completely the same as in Ref. \cite{SKTK2019}. With the same PNS models considered in this study, the gravitational wave frequencies with Cowling approximation have already been determined in Ref. \cite{ST2020b}. In Fig. \ref{fig:ft}, we show the time evolution of the frequency for several QNMs calculated without and with the Cowling approximation, where the filled marks with dashed lines (the open marks with dotted lines) are the results without (with) the Cowling approximation. From this figure, one can observe that the behavior of the frequencies obtained with the Cowling approximation is qualitatively the same way as in that without the Cowling approximation. In practice, the phenomena of avoided crossing can be seen in both of the time evolutions with and without the Cowling approximation, i.e., between the $f$- and $p_1$-modes at $T_{\rm pb}\simeq 0.2$ sec and between the $f$- and $g_1$-modes at $T_{\rm pb}\simeq 0.35$ sec.

%%%%%%%%%%%%%%%%%%%%%%%%%%%%%%%%%%%
% Figure 2
%%%%%%%%%%%%%%%%%%%%%%%%%%%%%%%%%%%
%%%%%%%%%%%%%%%%%%%%%%%%%%%%%%%%%%%

\begin{figure}[tbp]
\begin{center}
\includegraphics[scale=0.5]{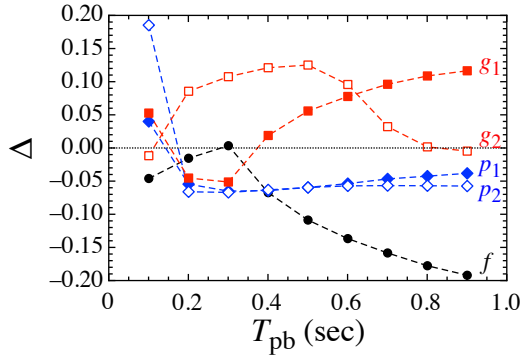} 
\end{center}
\caption{%%
Relative deviation of the frequency with the Cowling approximation from that without the Cowling approximation, calculated with Eq.~(\ref{eq:relative}).
}%%
\label{fig:Delta}
\end{figure}

In order to clearly see the accuracy of the Cowling approximation, we calculate the relative deviation between the frequencies obtained with and without the Cowling approximation as
\begin{equation}
  \Delta \equiv \frac{f_{\rm full} - f_{\rm Cowling}}{f_{\rm full}}, \label{eq:relative}
\end{equation}
where $f_{\rm full}$ and $f_{\rm Cowling}$ denote the frequencies determined without and with the Cowling approximation, respectively. In Fig. \ref{fig:Delta}, we show the value of $\Delta$ as a function of $T_{\rm pb}$, where the circle, filled-diamond, open-diamond, filled-square, and open-square correspond to the results for the $f$-, $p_1$-, $p_2$-, $g_1$-, and $g_2$-modes, respectively. From this figure, we find that the frequencies with the Cowling approximation can totally be determined within $\sim 20\%$ accuracy. In particular, the $p_i$-modes
except for $p_2$-mode at $T_{\rm pb}=0.1$ sec are within $\sim 7\%$ accuracy, the $g_i$-modes are determined within $\sim 13\%$ accuracy, while the $f$-mode in early phase ($T_{\rm pb}\lsim 0.4$ sec) are determined within $\sim 5\%$ accuracy. We also find that the $f$- and $p_i$-modes ($g_i$-modes) frequencies are basically overestimated (underestimated) with the Cowling approximation. Furthermore, in Fig.~\ref{fig:Delta} one can observe that the deviation in the $f$-mode frequency increases with time. Unfortunately, we cannot identify the reason of this result, but at least this result may be independent from the increase of the PNS  compactness, $M_{\rm PNS}/R_{\rm PNS}$, because the accuracy of the Cowling approximation for cold neutron stars becomes better as the stellar compactness increases \cite{YK1997}.

%%%%%%%%%%%%%%%%%%%%%%%%%%%%%%%%%%%
% Figure 3
%%%%%%%%%%%%%%%%%%%%%%%%%%%%%%%%%%%
%%%%%%%%%%%%%%%%%%%%%%%%%%%%%%%%%%%

Now, we try to compare our results with that in the previous studies \cite{MRBV2018,TCPOF2019a}, where some terms of the metric perturbations are taken into account. In Ref.~\cite{MRBV2018}, only the perturbation of the lapse function ($\delta\alpha$) is taken into account, and the $f$-mode frequency becomes higher than that with the Cowling approximation. On the other hand, in Ref.~\cite{TCPOF2019a}, the perturbations of the lapse function and conformal factor ($\delta\psi$) are taken into account, and the resultant frequency becomes lower than that with the Cowling approximation. In order to confirm the results of the previous study, we also calculate the frequencies by taking into account only $\delta\alpha$ or $\delta\alpha$ and $\delta\psi$. For this calculation, we adopt the open code, {\small GREAT}, provided in Ref. \cite{TCPOF2019a} by appropriately transforming the coordinate, because the metric in our formalism shown in Appendix \ref{sec:appendix_1} is given by
\begin{equation}
  ds^2 = -e^{2\Phi} dt^2 + e^{2\Lambda}dr^2 + r^2\left(d\theta^2 + \sin^2\theta d\phi^2\right),  \label{eq:metricsph} 
\end{equation}
where $\Phi$ and $\Lambda$ are metric functions with respect to only the circumference radius $r$, while {\small GREAT} is  prepared with the isotropic coordinate as 
\begin{equation}
  ds^2 = -\alpha^2 dt^2 + \psi^4\left[dr_{\rm iso}^2 + r_{\rm iso}^2\left(d\theta^2 + \sin^2\theta d\phi^2\right)\right],  \label{eq:metriciso} 
\end{equation}
where $\alpha$, $\psi$, and $r_{\rm iso}$ are the lapse function, conformal factor, and isotropic radial coordinate, respectively. So, as in Appendix B of Ref. \cite{Marek06} the transformation should be done through
\begin{gather}
  \alpha(r_{\rm iso}) = e^{\Phi(r)}, \label{eq:lapse} \\
  r_{\rm iso}\psi^2(r_{\rm iso}) = r, \label{eq:psi} \\
  \frac{d \ln r_{\rm iso}}{dr} = \frac{1}{r}\left(1-\frac{2m(r)}{r}\right)^{-1/2}, \label{eq:riso}
\end{gather}
where $\Phi$ and $m$ is phenomenological general relativistic potential and gravitational mass used in the simulation, respectively. We remark that since $\psi$ is set to be $1$ in the models with phenomenological general relativistic potential in the previous studies \cite{MRBV2018,TCPOF2019a}, we also show the result for the case with $\psi=1$ in Appendix \ref{sec:appendix_5}. In addition, the boundary condition at the PNS surface should be changed to that the Lagrangian perturbation of the pressure should be zero as in Ref. \cite{MRBV2018}, because the outer boundary condition in {\small GREAT} is imposed that the Lagrangian perturbation of radial coordinate of $r_{\rm iso}$ should be zero by default. Namely, we impose the boundary condition of 
\begin{equation}
    \rho h \alpha^{-2}\psi^4  \sigma^2 \eta_{\bot} 
    +\rho h\left(
     \frac{\delta \psi} {\psi}
    -\frac{\delta \alpha} {\alpha}
    \right)+ \eta_r\partial_r p  = 0  \label{eq:boundary}
\end{equation}
at the PNS surface 
with the notation in {\small GREAT}, where $\sigma$ is the eigenvalue, i.e., $\omega$ in our notation, $\eta_r$ and $\eta_\perp$ are variables associated with the Lagrangian displacement in the radial and angular directions, and $\rho,p,h$ are the density, pressure, and specific enthalpy, respectively  (See Ref. \cite{TCPOF2019a} for the meaning of the variables in detail).

%%%%%%%%%%%%%%%%%%%%%%%%%%%%%%%%%%%
% Figure 4
%%%%%%%%%%%%%%%%%%%%%%%%%%%%%%%%%%%
\begin{figure}[tbp]
\begin{center}
\includegraphics[scale=0.5]{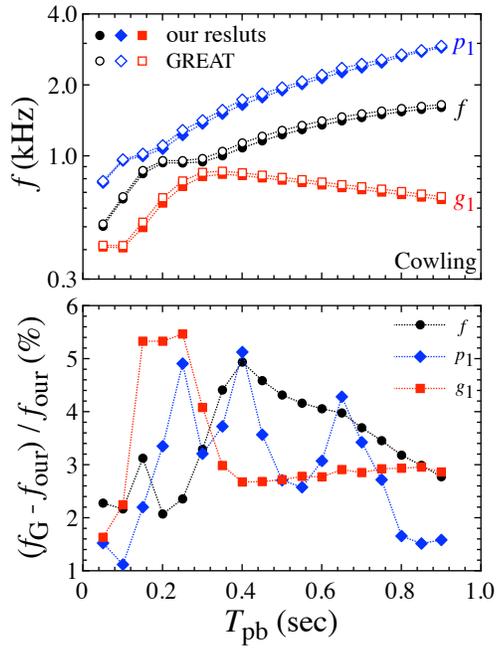} 
\end{center}
\caption{%%
Comparison between the frequencies with Cowling approximation calculated by us and with {\small GREAT}. In the top panel, the filled and open marks correspond to the results by us and with {\small GREAT}, respectively, where the circles, diamonds, and squares denote the $f$-, $p_1$-, and $g_1$-mode frequencies. In the bottom panel, we show the relative deviation in both frequencies. 
}%%
\label{fig:comp-Cow}
\end{figure}
%%%%%%%%%%%%%%%%%%%%%%%%%%%%%%%%%%%

%%%%%%%%%%%%%%%%%%%%%%%%%%%%%%%%%%%
% Figure 5
%%%%%%%%%%%%%%%%%%%%%%%%%%%%%%%%%%%
\begin{figure}[tbp]
\begin{center}
\includegraphics[scale=0.5]{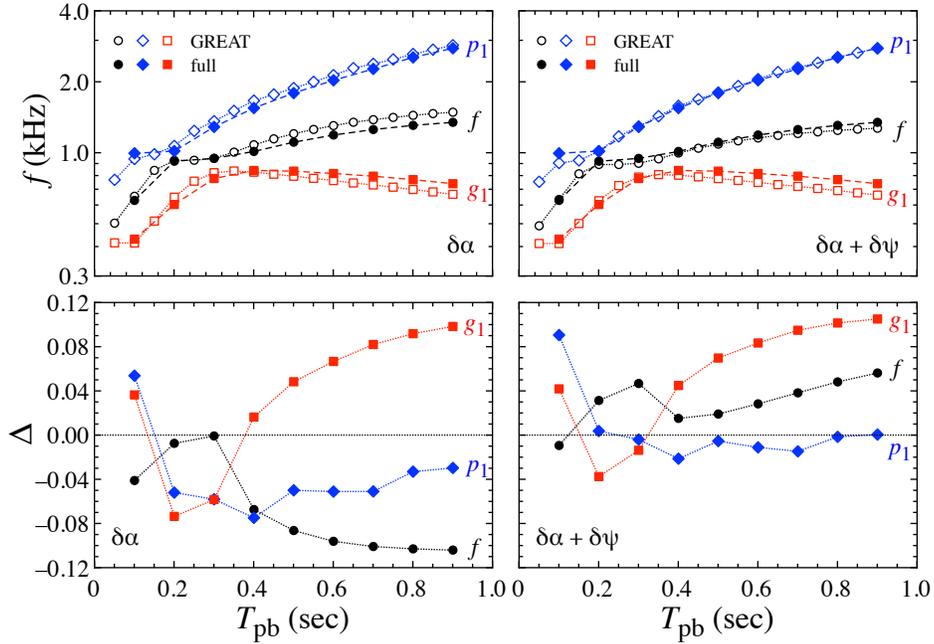} 
\end{center}
\caption{%%
The frequencies obtained with {\small GREAT} (open marks) are compared to those calculated without the Cowling approximation (filled marks) in the top panels, where the left panel correspond to the case with $\delta\alpha$, while the right panel is with $\delta\alpha$ and $\delta\psi$. The bottom panels denote the relative deviation, $\Delta$, from the frequencies without the Cowling approximation, where $\Delta$ is calculated in a similar way to Eq. (\ref{eq:relative}) but $f_{\rm Cowling}$ in the equation is replaced the frequencies with $\delta\alpha$ or with $\delta\alpha$ and $\delta\psi$.
}%%
\label{fig:comp-daldap}
\end{figure}
%%%%%%%%%%%%%%%%%%%%%%%%%%%%%%%%%%%

\begin{figure}[tbp]
\begin{center}
\includegraphics[scale=0.6]{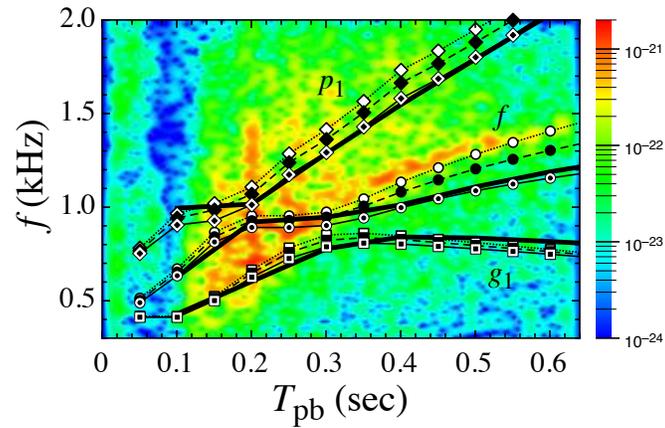} 
\end{center}
\caption{%%
Time evolution of the $f$-, $p_1$-, and $g_1$-mode frequencies calculated with the Cowling approximation (open marks), with $\delta \alpha$ (filled marks), with $\delta\alpha$ and $\delta\psi$ (double marks), and will full perturbations (thick-solid line) are shown in the gravitational wave spectrogram obtained in Ref. \cite{ST2020b}. 
}%%
\label{fig:2D}
\end{figure}

First, in order to check whether we can felicitously reproduce the same physical results with {\small GREAT}, even though the choice of coordinates and boundary conditions differ, we compare the frequencies in the Cowling approximation calculated with our code and with {\small GREAT}. This is because the outer boundary condition is set to be $\eta_r=0$ as default in {\small GREAT}, while that in our problem should be Eq. (\ref{eq:boundary}), and the coordinate system in {\small GREAT} is also  different from ours. The results are shown in Fig. \ref{fig:comp-Cow}, where the top panel shows the $f$-, $p_1$-, and $g_1$-mode frequencies calculated with our code (filled marks) and with {\small GREAT} (open marks), while in the bottom panel we show the relative deviation in both frequencies. From this figure, we conclude that we can felicitously handle {\small GREAT} even in our problem. We remark that the difference in Fig. \ref{fig:comp-Cow} comes from the coordinate transformation. That is, to adopt {\small GREAT}, we have to transform our radial coordinate ($r$) to that in {\small GREAT} ($r_{\rm iso}$) with discretized background data. Then, we compare the frequencies with $\delta\alpha$  (allowing for perturbations to $\alpha$) and with $\delta\alpha$ and $\delta\psi$ (allowing for perturbations to $\alpha$ and $\psi$) to that without the Cowling approximation (with full metric perturbations). In Fig. \ref{fig:comp-daldap} we show the corresponding frequencies in the top panels (the left panel with $\delta\alpha$ and the right panel with $\delta\alpha$ and $\delta\psi$), where the open and filled  marks denote the frequencies calculated with {\small GREAT} and with the full metric perturbations, respectively. In the bottom panels in Fig. \ref{fig:comp-daldap}, we show the relative deviation of the frequencies with $\delta\alpha$ (left panel) and with $\delta\alpha$ and $\delta\psi$ (right panel) from the frequencies with the full metric perturbations, which is calculated in a similar way to Eq. (\ref{eq:relative}), but $f_{\rm Cowling}$ in the equation should be replaced to the frequencies with $\delta\alpha$ or with $\delta\alpha$ and $\delta\psi$. From this figure together with Fig. \ref{fig:Delta}, we find that the $f$-mode frequencies calculated with $\delta\alpha$ or with $\delta\alpha$ and $\delta\psi$ become more accurate than that with the Cowling approximation, but the $g$-mode frequencies are not so improved by taking into account some of the metric perturbations. In Fig. \ref{fig:2D}, we show the $f$-, $p_1$-, and $g_1$-mode frequencies calculated with the Cowling approximation, with $\delta\alpha$, with $\delta\alpha$ and $\delta\psi$, and with full perturbations on the gravitational wave spectrogram obtained in Ref. \cite{ST2020b}. Anyway, our results, where the $f$-mode frequencies with the Cowling approximation (e.g., $657$, $1229$, $1605$ Hz at $T_{\rm pb} =0.1$, 0.5, and 0.9 sec.) become larger than those with $\delta\alpha$ (e.g., $653$, $1204$, $1486$ Hz at $T_{\rm pb} =0.1$, 0.5, and 0.9 sec.) are a similar tendency to the results shown in Ref. \cite{TCPOF2019a} (see also Fig.~\ref{fig:2D-spe} in Appendix \ref{sec:appendix_5}).

Though the aim of this study is to show the systematic difference between the Cowling approximation and the full perturbation (see Fig. \ref{fig:Delta}) and not to provide a better fit of the frequency with the numerical simulation,
we briefly comment on a gap of the frequency of the ramp up signals between the linear analysis and the numerical simulation.
As shown in Ref. \cite{ST2020b}, the signals in the numerical simulation are a little higher than the $f$-mode frequency calculated with the Cowling approximation. In addition, we show in this study that the $f$-mode frequency with the Cowling approximation is overestimated, which leads to the result that the difference between the $f$-mode frequency without the Cowling approximation and the ramp up signals in the numerical simulation, becomes larger (see Fig. \ref{fig:2D}). This difference could come from the treatment of gravity in the numerical simulations, i.e., the numerical simulations in Refs. \cite{ST2020b} has employed gravity with the phenomenological general relativistic effect, while the linear theory is done in the framework of full general relativity.  In order to validate this hypothesis, further study using models of full general relativity is necessary somewhere in future. To summarize the current status, we demonstrate how the treatment of gravity in the linear analysis, affects the frequency in Appendix \ref{sec:appendix_5}.

%%%%%%%%%%%%%%%%%%%%%%%%%%%%%%%%%%%
% Figure 6
%%%%%%%%%%%%%%%%%%%%%%%%%%%%%%%%%%%
%%%%%%%%%%%%%%%%%%%%%%%%%%%%%%%%%%%

\begin{figure}[tbp]
\begin{center}
\includegraphics[scale=0.5]{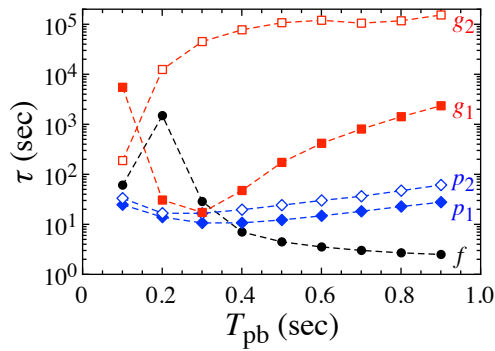} 
\end{center}
\caption{%%
Time evolution of the damping time for the $f$-, $p_1$-, $p_2$-, $g_1$-, and $g_2$-modes.
}%%
\label{fig:taut}
\end{figure}

Finally, as an advantage by taking into account the metric perturbations without the Cowling approximation, now we can also discuss the damping time (or damping rate) of QNMs. In Fig. \ref{fig:taut}, we show the resultant damping time is shown as a function of $T_{\rm pb}$, where the meaning of the different marks is the same as in Fig. \ref{fig:Delta}. These behaviors seem to be qualitatively similar to that shown in Ref. \cite{FMP2003,Camelio17}, even though the PNS models in Ref. \cite{FMP2003,Camelio17} are assumed that the baryonic mass is constant during PNS evolution for $T_{\rm pb} \ge 0.2$ sec, i.e., the mass accretion effect is neglected, which leads to that the gravitational mass of PNS monotonically decreases as time.

As shown in Fig. \ref{fig:taut}, we find that the damping time of each mode is longer than $1$ sec, which is long enough compared to the time interval in a short-time Fourier transform for data analysis (e.g., Ref.  \cite{Murphy09}), which is usually set to be $\sim 0.02$ sec. During the interval, we can safely assume that the waveform is a plane wave like $\sin(\omega t)$ or $\cos(\omega t)$, where the damping factor of $\exp(-t/\tau)$ can be ignored. The actual limit on the assumption would come from the timescale in which the background changes or too many seed perturbations that are imposed on the PNS.

In Fig. \ref{fig:taut}, we also find that unlike the oscillation frequency, the damping time can cross between different modes. As a result, we also find that the damping time of $g_1$- and $p_i$-modes can be smaller than that of the $f$-mode before the avoided crossing in the frequency between the $f$- and $g_1$-modes, i.e., before $T_{\rm pb}\simeq 0.35$, while the damping time for QNMs after $T_{\rm pb}\simeq 0.35$ is more or less similar to that in cold neutron stars. In particular, we find that the damping time for $g_2$-mode is quite long, i.e., this mode hardly damps out, with almost constant frequency as shown in Fig. \ref{fig:ft} (and in Fig. 2 in Ref. \cite{ST2020b}). So, the $g_2$-mode in the PNSs could effectively excite the gravitational waves, although it strongly depends on the oscillation energy in such a mode, which is still unknown.  In fact, in a similar way to Refs.~\cite{AK1996,AK1998}, the effective amplitude, $h_{\rm eff}$, for the oscillation mode with the frequency $f$, damping time $\tau$, the released energy $E$, and the distance between the source and the Earth $D$, is estimated as
\begin{equation}
    h_{\rm eff} \simeq 
    h \sqrt{f\tau} \simeq \frac{1}{\sqrt{2}\pi D}\sqrt{\frac{E}{f}},
\end{equation}
which leads to the effective amplitude of the $f$- and $g_2$-modes as 
\begin{gather}
  h_{\rm eff}^{(f)} \simeq 1.54\times 10^{-21} \left(\frac{E_f}{10^{-8}M_\odot}\right)^{1/2}\left(\frac{1\ {\rm kHz}}{f_f}\right)^{1/2}\left(\frac{10\ {\rm kpc}}{D}\right), \\
  h_{\rm eff}^{(g_2)} \simeq 2.43\times 10^{-22} \left(\frac{E_{g_2}}{10^{-10}M_\odot}\right)^{1/2}\left(\frac{0.4\ {\rm kHz}}{f_{g_2}}\right)^{1/2}\left(\frac{10\ {\rm kpc}}{D}\right), 
\end{gather}
where $E_i$ and $f_i$ (for $i=f$ or $g_2$) denote the energy released by each mode and the oscillation frequency, respectively. Here, we assume the typical values of $E_f$ and $E_{g_2}$ so that the effective amplitude becomes $h_{\rm eff}^{(f)}\sim 10^{-21}$ and $h_{\rm eff}^{(g_2)}\sim 10^{-22}$ as Fig. \ref{fig:2D}. But, if one considers the case with $E_f=E_{g_2}$, $h_{\rm eff}^{(g_2)}$ becomes comparable to $h_{\rm eff}^{(f)}$. 

%%%%%%%%%%%%%%%%%%%%%%%%%%%%%%%%%%%
% Figure 7
%%%%%%%%%%%%%%%%%%%%%%%%%%%%%%%%%%%
%%%%%%%%%%%%%%%%%%%%%%%%%%%%%%%%%%%

%%%%%%%%%%%%%%%%%%%%%%%%%%%%%%%%%%%%%%%%%%%%%%%%
\section{Conclusion}
\label{sec:Conclusion}
%%%%%%%%%%%%%%%%%%%%%%%%%%%%%%%%%%%%%%%%%%%%%%%%

In order to check the accuracy of the relativistic Cowling approximation for the gravitational waves from the PNSs provided just after the supernova explosions, we calculated the frequencies with and without the Cowling approximation for the PNS models constructed with the 2D numerical simulation of core-collapse supernova. Then, we found that the behavior of the frequency evolution calculated with the Cowling approximation is qualitatively the same way as in that without the Cowling approximation, where one can observe the avoided crossing in the frequency evolution after core-bounce even without the Cowling approximation at the same time as that with the Cowling approximation. Thus, even with the Cowling approximation, one can discuss the behavior of the gravitational wave frequencies from the PNSs. We also showed that the frequencies with the Cowling approximation can be determined within $\sim 20\%$ accuracy, where the $p_i$-modes are basically determined well with $\sim 7\%$ accuracy. Note that the $f$- and $p_i$-mode frequencies with the Cowling approximation are typically overestimated. In addition, as an advantage by taking into account the metric perturbations, we showed the time evolution of the damping time for the several quasi-normal modes. The damping time is long enough compared to the typical time interval of short-Fourier transformation that often used in the analysis. This finding ideally guarantees the validity of the method unless the background changes rapidly or too many seed perturbations are imposed. In particular, we found that the damping time of the $g_2$-mode is quite long, while its frequency is almost constant in time. So, this mode may efficiently be excited in the gravitational waves in the PNSs.

%\newpage
%%%%%%%%%%%%%%%%%%%%%%%%%%%%%%%%%%%%%%%%%%%%%%%%
\acknowledgments
%%%%%%%%%%%%%%%%%%%%%%%%%%%%%%%%%%%%%%%%%%%%%%%%

This work is supported in part by Grant-in-Aid for Scientific Research (Nos.
JP17H01130, %Kotake-san Kiban A  (Takiwaki Co-PI)
JP17K14306, %Takiwaki-san Wakate B 
JP18H01212, % Yokoi Kinban B (Takiwaki Co-PI)
JP19KK0354 % Sotani 
) from Japan Society for the Promotion of Science (JSPS), 
Grant-in-Aid for Scientific Research on Innovative Areas (Nos.
JP17H06357, %Shingakujutu GWGEN Sokatsu
JP17H06364, %Shingakujutu GWGEN C01,  (Takiwaki Co-PI)
JP20H04753 % Sotani Shingakujutu-Koubo
) from the Ministry of Education, Science and Culture of Japan (MEXT).
This research has been also supported by MEXT as “Program for Promoting Researches on the Supercomputer Fugaku” (Toward a unified view of the universe: from large scale structures to planets) and JICFuS,
the National Institutes of Natural Sciences 
(NINS) program for cross-disciplinary
study (Grant Numbers 01321802 and 01311904) on Turbulence, Transport,
and Heating Dynamics in Laboratory and Solar/Astrophysical Plasmas:
"SoLaBo-X”.
Numerical computations were in part carried out on Cray XC50, PC cluster and analysis server at Center for Computational Astrophysics, National Astronomical Observatory of Japan.

\appendix
%%%%%%%%%%%%%%%%%%%%%%%%%%%%%%%%%%%%%%%%%%%%%%%%
\section{Perturbation equations inside the PNS}   % Appendix A
\label{sec:appendix_1}
%%%%%%%%%%%%%%%%%%%%%%%%%%%%%%%%%%%%%%%%%%%%%%%%

The metric for spherically symmetric background is expressed by Eq. (\ref{eq:metricsph}). Additionally, $\Lambda$ is directly associated with the mass function, $m(r)$, as $e^{-2\Lambda}=1-2m/r$. Then, for the case of the polar-type oscillations, the metic perturbations, $h_{\mu\nu}$, are given by
\begin{equation}
 h_{\mu\nu} =
 \left(
 \begin{array}{cccc}
 r^\ell e^{2\Phi} H  &  i\omega r^{\ell+1} H_1 & 0 & 0 \\
 i\omega r^{\ell+1} H_1 & r^{\ell} e^{2\Lambda} H & 0 & 0 \\
 0& 0 & r^{\ell+2} K & 0 \\
 0 & 0 & 0 & r^{\ell+2} K \sin^2\theta \\
 \end{array}
 \right) e^{i\omega t}Y_{\ell k}\,, \label{eq.A1}
 \end{equation}
where $H$, $H_1$, and $K$ are functions of $r$ \cite{LD83,STM2001}, while the Lagrangian displacement, $\xi^i$, of the fluid element is given by 
\begin{gather}
  \xi^r(t,r,\theta,\phi) = r^{\ell-1} e^\Lambda W e^{i\omega t} Y_{\ell k}, \\
  \xi^\theta(t,r,\theta,\phi) = -r^{\ell-2} e^\Lambda V e^{i\omega t} \partial_\theta Y_{\ell k}, \\
  \xi^\phi(t,r,\theta,\phi) = -r^{\ell-2} e^\Lambda V e^{i\omega t} \sin^{-2}\theta \,\partial_\phi Y_{\ell k},
\end{gather}
where $W$ and $V$ are functions of $r$. We remark that since the $(i,2)$- and $(i,3)$-components with $i=0,1$ in Eq. (\ref{eq.A1}) correspond to the axial-type oscillations, we omit these terms in this study. Here, $\omega$ corresponds to a complex eigenvalue. In addition to the perturbation variables, $H$, $H_1$, $K$, $W$, and $V$, we introduce an auxiliary variable, $X$, defined as
\begin{equation}
  X\equiv \omega^2 (p+\varepsilon) e^{-\Phi} V - \frac{e^{\Phi-\Lambda}}{r}\frac{dP}{dr}W -\frac{1}{2}(p+\varepsilon)e^{\Phi}H.  \label{eq:XX}
\end{equation}
This is because the Lagrangian perturbation of pressure, $\Delta p$, is given by $\Delta p = -r^\ell e^{-\Phi}XY_{\ell k}$, which make the boundary condition at the PNS surface easy (see Appendix \ref{sec:appendix_3}).
With these variables, the perturbation equations obtained from the linearized Einstein equations are expressed as
\begin{gather}
  \frac{dH_1}{dr} = -\frac{1}{r}\left[\ell+1+\frac{2m}{r}e^{2\Lambda}+4\pi r^2(p-\varepsilon)e^{2\Lambda}\right]H_1
      + \frac{1}{r}e^{2\Lambda}\left[H+K+16\pi(p+\varepsilon)V\right], \label{eq:dH1} \\
  \frac{dK}{dr} = \frac{\ell(\ell+1)}{2r}H_1 + \frac{H}{r} - \left[\frac{\ell+1}{r} - \frac{d\Phi}{dr}\right]K 
      + \frac{8\pi}{r}(p+\varepsilon)e^{\Lambda}W,  \label{eq:dK} \\
  \frac{dW}{dr} = -\frac{\ell+1}{r}W + re^\Lambda\left[\frac{1}{\Gamma p}e^{-\Phi}X - \frac{\ell(\ell+1)}{r^2}V
      - \frac{H}{2} - K \right],  \label{eq:dW} \\
  \frac{dX}{dr} = -\frac{\ell}{r}X + (p+\varepsilon)e^{\Phi} \bigg[\frac{1}{2}\left(\frac{d\Phi}{dr} - \frac{1}{r}\right)H 
     - \frac{1}{2}\left(\omega^2 r e^{-2\Phi} + \frac{\ell(\ell+1)}{2r}\right)H_1 + \left(\frac{1}{2r} - \frac{3}{2}\frac{d\Phi}{dr}\right)K \nonumber \\
     - \frac{\ell(\ell+1)}{r^2}\frac{d\Phi}{dr}V - \frac{1}{r}\left(\omega^2 e^{-2\Phi + \Lambda} + 4\pi(p+\varepsilon)e^{\Lambda}
     - r^2\left\{\frac{d}{dr}\left(\frac{1}{r^2}e^{-\Lambda}\frac{d\Phi}{dr}\right)\right\}\right)W \bigg],  \label{eq:dX}
\end{gather}
together with two algebraic equations for $H$ and $V$ such as
\begin{gather}
  \left[1 - \frac{3m}{r} - \frac{\ell(\ell+1)}{2} - 4\pi r^2 p\right]H - 8\pi r^2 e^{-\Phi}X 
     + r^2 e^{-2\Lambda} \left[\omega^2 e^{-2\Phi} - \frac{\ell(\ell+1)}{2r}\frac{d\Phi}{dr}\right]H_1 \nonumber \\
     - \left[1 + \omega^2 r^2 e^{-2\Phi} - \frac{\ell(\ell+1)}{2} - (r - 3m - 4\pi r^3 p)\frac{d\Phi}{dr}\right]K = 0, \label{eq:const-H} \\
  V= \frac{e^{2\Phi}}{\omega^2(p+\varepsilon)}\left[e^{-\Phi}X + \frac{1}{r}\frac{dP}{dr}e^{-\Lambda}W + \frac{1}{2}(p+\varepsilon)H\right]. \label{eq:const-V}
\end{gather}
In Eq. (\ref{eq:dW}), $\Gamma$ denotes the adiabatic index for the background stellar model, which is associated with the sound velocity, $c_s$, as
\begin{equation}
  \Gamma = \frac{p + \varepsilon}{p} c_s^2. 
\end{equation}
So, inside the PNS, one has to integrate four  1st order differential equations, i.e., Eqs.  (\ref{eq:dH1}) - (\ref{eq:dX}), together with two constraint equations, i.e., Eqs. (\ref{eq:const-H}) and (\ref{eq:const-V}). We remark that Eq. (\ref{eq:const-H}) comes from the linearlized Einstein equations, while Eq. (\ref{eq:const-V}) just corresponds to Eq. (\ref{eq:XX}).

%%%%%%%%%%%%%%%%%%%%%%%%%%%%%%%%%%%%%%%%%%%%%%%%
\section{Perturbation equations outside the PNS}   % Appendix B
\label{sec:appendix_2}
%%%%%%%%%%%%%%%%%%%%%%%%%%%%%%%%%%%%%%%%%%%%%%%%

The matter density beyond the surface of the PNS is nonzero, where the PNS is considered as the region inside the specific surface density in this study. Even so, since the density outside the PNS is much smaller than that inside the PNS, for simplicity we assume that the exterior region of the PNS would be vacuum, where we neglect the gravitational waves generated in the exterior region of the PNS. Then, the perturbation equations become the so-called Zerilli equations. In practice, by setting $m=M_{\rm PNS}$ and $X=W=V=0$, and by replacing $H_1$ and $K$ with new variables, $Z$ and $dZ/dr_*$, defined as
\begin{gather}
  r^\ell K = \frac{\lambda(\lambda+1)r^2 + 3\lambda r M_{\rm PNS} 
     + 6M_{\rm PNS}^2}{r^2(\lambda r + 3M_{\rm PNS})}Z + \frac{dZ}{dr_*}, \\
  r^{\ell+1}H_1 = \frac{\lambda r^2 - 3\lambda r M_{\rm PNS} - 3M_{\rm PNS}^2}{(r-2M_{\rm PNS})(\lambda r + 3M_{\rm PNS})}Z 
     + \frac{r^2}{r-2M_{\rm PNS}}\frac{dZ}{dr_*},
\end{gather} 
one can reduce the Zerilli equations, where $Z$ is the Zerilli function, $r_*$ is the tortoise coordinate defined by
\begin{equation}
  r_* \equiv r + 2M_{\rm PNS} \ln\left(\frac{r}{2M_{\rm PNS}} -1 \right),
\end{equation}
and $\lambda\equiv \ell(\ell+1)/2 -1$. That is, $\partial_r = e^{\Lambda -\Phi}\partial_{r_*}$.

%%%%%%%%%%%%%%%%%%%%%%%%%%%%%%%%%%%%%%%%%%%%%%%%
\section{Boundary conditions}   % Appendix C
\label{sec:appendix_3}
%%%%%%%%%%%%%%%%%%%%%%%%%%%%%%%%%%%%%%%%%%%%%%%%

To solve the perturbation equations given in Appendix \ref{sec:appendix_2} and \ref{sec:appendix_3}, one has to impose appropriate boundary conditions at the stellar center, the stellar surface, and the spatial infinity. The boundary condition at the stellar center is the regularity condition for all the perturbation variables, which reduces to the following relations at the stellar center \cite{LD83,STM2001}: 
\begin{gather}
  H_1 = \frac{1}{\ell(\ell+1)}\left[2\ell K - 16\pi(p+\varepsilon)W\right],  \\ 
  X = (p+\varepsilon)e^{\Phi}\left[-\frac{1}{2}K + \left(\frac{4\pi}{3}\varepsilon + 4\pi p 
      - \frac{\omega^2}{\ell}e^{-2\Phi}\right)W\right], \\
  H = K, \\
  V = -\frac{1}{\ell}W.
\end{gather}
The boundary condition imposed at the stellar surface is that the Lagrangian perturbation of pressure should be zero, which corresponds to $X=0$ at the stellar surface, while the metric perturbations should be continuous. The boundary condition at the spatial infinity is that only the outgoing gravitational waves exist.

%%%%%%%%%%%%%%%%%%%%%%%%%%%%%%%%%%%%%%%%%%%%%%%%
\section{How to determine the quasi-normal modes}   % Appendix D
\label{sec:appendix_4}
%%%%%%%%%%%%%%%%%%%%%%%%%%%%%%%%%%%%%%%%%%%%%%%%

In order to determine the quasi-normal modes (or in order to deal with the outer boundary condition at the spatial infinity), we follow the method developed in Ref. \cite{LNS93}. First, the Zerilli function ($Z$) is changed to the Regge-Wheeler function ($Q$), because the effective potential in the Regge-Wheeler equation is simpler than that in the Zerilli equation. According to Ref. \cite{Chandrasekhar83}, this transformation can be done via 
\begin{gather}
  (\kappa + 2i \omega\beta)Z = (\kappa + 2\beta^2 f(r))Q + 2\beta\frac{dQ}{dr_*}, \\
  (\kappa - 2i \omega\beta)Q = (\kappa + 2\beta^2f(r))Z - 2\beta\frac{dZ}{dr_*},
\end{gather}
where $\beta\equiv 6M_{\rm PNS}$, $\kappa\equiv 4\lambda(\lambda+1)$, and 
\begin{equation}
   f(r) \equiv \frac{r-2M_{\rm PNS}}{2r^2(\lambda r + 3M_{\rm PNS})}. 
\end{equation}
Then, by expanding the variable $Q$ at a finite radius, $r=r_a$, as
\begin{equation}
  Q(r) = \left(\frac{r}{2M_{\rm PNS}} - 1\right)^{-2i \omega M_{\rm PNS}} e^{-i\omega r}
     \sum_{n=0}^{\infty}a_n\left(1 - \frac{r_a}{r}\right)^n, 
\end{equation}
where any terms proportional to $\exp(i\omega r)$, i.e., ingoing wave, are removed far from the PNS, and by substituting this expression into the Regge-Wheeler equation, one can obtain a four-term recurrence relation with respect to $a_n$ for $n\ge 2$, such as
\begin{equation}
  \alpha_n a_{n+1} + \beta_n a_n + \gamma_n a_{n-1} + \delta_n a_{n-2} = 0,   \label{eq:4term}
\end{equation}
where the coefficients are given by 
\begin{gather}
  \alpha_n = \left(1-\frac{2M_{\rm PNS}}{r_a}\right)n(n+1),  \label{eq:alpha0} \\
  \beta_n = -2\left[i\omega r_a + \left(1- \frac{3M_{\rm PNS}}{r_a}\right)n\right]n,  \label{eq:beta0} \\
  \gamma_n = \left(1 - \frac{6M_{\rm PNS}}{r_a}\right)n(n-1) + \frac{6M_{\rm PNS}}{r_a} - \ell(\ell+1),  \label{eq:gamma0}  \\
  \delta_n = \frac{2M_{\rm PNS}}{r_a}(n-3)(n+1).  \label{eq:delta0}
\end{gather}
In particular, the values of $a_{-1}$, $a_0$, and $a_1$ are determined with the values of $Q$ and $dQ/dr$ at $r=r_a$ as
\begin{gather}
  a_{-1} = 0, \ \ 
  a_0 = \frac{Q(r_a)}{\chi(r_a)}, \\
  a_1 = \frac{r_a}{\chi(r_a)}\left[\frac{dQ}{dr}\bigg|_{r_a} + \frac{i\omega r_a}{r_a - 2M_{\rm PNS}}Q(r_a)\right], 
\end{gather}
where 
\begin{equation}
  \chi(r) \equiv \left(\frac{r}{2M_{\rm PNS}} - 1\right)^{-2i \omega M_{\rm PNS}}e^{-i\omega r}.
\end{equation}
Moreover, by defining new coefficients, $\hat{\alpha}_n$, $\hat{\beta}_n$, and $\hat{\gamma}_n$, with Eqs. (\ref{eq:alpha0}) -- (\ref{eq:delta0}) as $\hat{\alpha}_1=\alpha_1$, $\hat{\beta}_1=\beta_1$, $\hat{\gamma}_1=\gamma_1$, and for $n\ge 2$
\begin{gather}
  \hat{\alpha}_n = \alpha_n, \\
  \hat{\beta}_n = \beta_n - \frac{\hat{\alpha}_{n-1}\delta_n}{\hat{\gamma}_{n-1}}, \\
  \hat{\gamma}_n = \gamma_n - \frac{\hat{\beta}_{n-1}\delta_n}{\hat{\gamma}_{n-1}},
\end{gather}
the four-term recurrence relation given by Eq. (\ref{eq:4term}) is reduced to a three-term recurrence relation as
\begin{equation}
  \hat{\alpha}_n a_{n+1} + \hat{\beta}_n a_n + \hat{\gamma}_n a_{n-1} = 0.
\end{equation}
With this recurrence relation, one can derive the form of continued fraction composed of $\hat{\alpha}_n$, $\hat{\beta}_n$, and $\hat{\gamma}_n$, such as
\begin{equation}
  \frac{a_1}{a_0} = \frac{-\hat{\gamma}_1}{\hat{\beta}_1-}\frac{\hat{\alpha}_1\hat{\gamma}_2}{\hat{\beta}_2-}
      \frac{\hat{\alpha}_2\hat{\gamma}_3}{\hat{\beta}_3-}\cdots, 
\end{equation}
or 
\begin{equation}
  0= \hat{\beta}_0 - \frac{\hat{\alpha}_0\hat{\gamma}_1}{\hat{\beta}_1-}\frac{\hat{\alpha}_1\hat{\gamma}_2}{\hat{\beta}_2-}
      \frac{\hat{\alpha}_2\hat{\gamma}_3}{\hat{\beta}_3-}\cdots \equiv {\cal F}(\omega), 
\end{equation}
where $\hat{\beta}_0 = a_1/a_0$ and $\hat{\alpha}_0=-1$. Finally, the eigenvalue, $\omega$, can be determined by solving the equation of ${\cal F}(\omega)=0$, where we adopt $n_{\rm max}=200000$ which is the maximum value of $n$.

%%%%%%%%%%%%%%%%%%%%%%%%%%%%%%%%%%%%%%%%%%%%%%%%
\section{For the case with $\psi=1$}   % Appendix E
\label{sec:appendix_5}
%%%%%%%%%%%%%%%%%%%%%%%%%%%%%%%%%%%%%%%%%%%%%%%%

\begin{figure}[tbp]
\begin{center}
\includegraphics[scale=0.5]{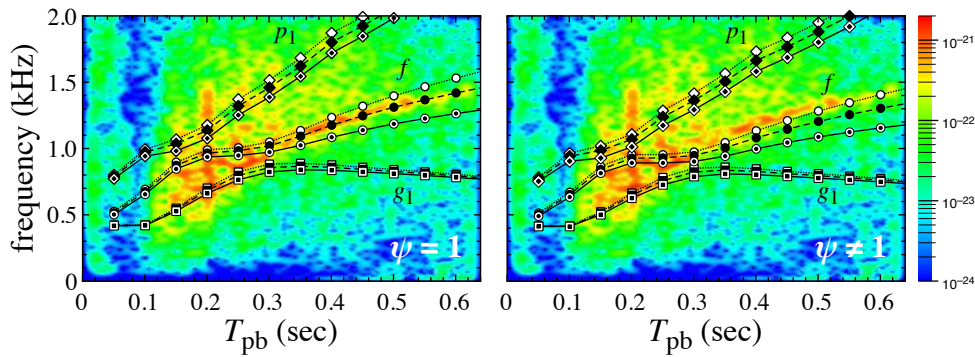} 
\end{center}
\caption{%%
The $f$-, $p_1$-, and $g_1$-mode frequencies calculated with {\small GREAT} are shown in the gravitational wave spectrogram obtained from the data of numerical simulation \cite{ST2020b}, where the left panel corresponds to the frequencies calculated by setting $\psi=1$, while the right panel corresponds to those with $\psi$ appropriately calculated through Eq. (\ref{eq:psi}), i.e., the same as Fig. \ref{fig:2D} without the result for full perturbations. The open, filled, and double marks denote the frequencies calculated with the Cowling approximation, with $\delta\alpha$, and with $\delta\alpha$ and $\delta\psi$, respectively. 
}%%
\label{fig:2D-spe}
\end{figure}

In this study, to adopt the open code, {\small GREAT}, consistently with our coordinate system, we transform the metric function together with the radial coordinate via Eqs. (\ref{eq:lapse}) - (\ref{eq:riso}). On the other hand, in the previous studies \cite{MRBV2018,TCPOF2019a}, the frequencies are calculated by setting $\psi=1$, where $r_{\rm iso}$ becomes equivalent to $r$. So, in this appendix, we  compare the frequencies with $\psi=1$ to those with $\psi$ calculated with Eq. (\ref{eq:psi}) just for reference. In Fig. \ref{fig:2D-spe} we show the $f$-, $p_1$-, and $g_1$-mode frequencies with $\psi=1$ in the left panel and those with $\psi$ calculated with Eq. (\ref{eq:psi}) in the right panel, where the background on each panel denotes the gravitational wave spectrogram obtained from the data of the corresponding numerical simulation. In the figure, open, filled, and double marks denote the frequencies with the Cowling approximation, with $\delta\alpha$, and with $\delta\alpha$ and $\delta\psi$, respectively.

From this figure, one can see that the frequencies calculated with $\psi=1$ are estimated larger than those with $\psi$ calculated with Eq. (\ref{eq:psi}). In addition, in both cases with $\psi=1$ and $\psi\ne 1$, the frequencies with $\delta\alpha$ become smaller than those with the Cowling approximation, and the frequencies with  $\delta\alpha$ and $\delta\psi$ become smaller than those with $\delta\alpha$. This tendency is consistent with the results shown in Ref. \cite{TCPOF2019a}. Among the models, the estimate with lapse perturbation in $\psi=1$ provides the best fit for the numerical simulation. We remark that this setting is closer to what has been done in the numerical simulation, i.e., the Newtonian simulation ($\psi=1$) with phenomenological gravitational potential ($\delta \alpha\neq 0$).

%%%%%%%%%%%%%%%%%%%%%%%%%%%%%%%%%%%
% Figure 8
%%%%%%%%%%%%%%%%%%%%%%%%%%%%%%%%%%%
%%%%%%%%%%%%%%%%%%%%%%%%%%%%%%%%%%%

Additionally, we try to compare our models with the models in the market. Table~\ref{tab:comp_sim} summarizes the setting and the corresponding  results with the linear analysis in literature and ours. All the background models in the table are performed in the phenomenological general relativistic potential as in Ref. \cite{Marek06}. We summarize the setup of linear analysis in the columns of the ``Setup of linear analysis". The detail is shown in the caption. In a private communication to the authors of Ref.~\cite{TCPOF2019a}, we know the settings of $\alpha=1+\Phi$ and  $\psi=1$ is used in T19m20. Since $1+\Phi $ should be identical to $\exp(\Phi)$ if $|\Phi| \ll 1$, we do not focus on the difference in the treatment of the lapse function. In the right half of the table is dedicated to describing the results in the specific settings, where ``high", ``just", and ``low" in each column respectively correspond to the case when the $f$-mode frequency obtained with the linear analysis is larger than, comparable to, and lower than the ramp up signals in the numerical simulation, where the judgment of "just" is simply by eye. Note that Ref.~\cite{TCPOF2019a} has used a different classification method. We call their $^2g_1$-mode as $f$-mode.

Basically, it is not easy to investigate the consistency of our results to the previous works, since the coordinate system, the boundary condition, the setup for background metric differ each other. However, we can find some features. The models with $\psi \neq 1$ or $\exp(\Lambda )\neq 1$, i.e., m2.9SGR and m2.9IGR, predict systematically lower frequency compared to that in the numerical simulations. That indicates the treatment of the background metric is important. On the other hand, among the models with $\psi=1$, the setups of M18m10 and m2.9INW are similar, but the results are not consistent:  M18m10 predicts lower frequencies compared to the simulation, while m2.9IGR  predicts higher frequencies than that. We need further investigation on this issue in the future. Despite that the boundary condition is different, T19m20 and m2.9INW show similar trends. The effect of the boundary condition on this issue may not be so large.

%%%%%%%%%%%%%%%%%%%%%%
%  Table 1
%%%%%%%%%%%%%%%%%%%%%%
\begin{table}
\centering
\caption{Summary of the setting and the results of the linear analysis in literature and ours. 
In the columns of the ``Setup of linear analysis", we summarize the setup. The ``coordinates" column refers to the coordinate system of the linear analysis: isotropic or Schwarzschild coordinates (spherical coordinates) are used. The ``boundary" column shows the choice of the boundary condition: ``PNS, $\Delta P=0$" means that the Lagrangian perturbation of the pressure should be zero at the PNS surface, while ``shock, $\eta_r=0$" means that the Lagrangian perturbation of radial coordinate should be zero at the shock radius. The ``lapse" and ``conformal factor" columns denote how to set the background $\alpha$ and $\psi$ or $\Lambda$ in Eq. \eqref{eq:metricsph} and Eq.~\eqref{eq:metriciso}.
In the columns, $\exp(\Lambda)\neq 1$ or $\psi \neq 1$ means the employment of the general relativistic effect.
In the right half of the table is dedicated to describing the results in the specific settings.
The relative frequency of the $f$-mode in the linear analysis with respect to that in the numerical simulation is written in each column.}
\begin{tabular}{c|cccc|cccc}
\hline
\hline
Models                                           & \multicolumn{4}{c|}{Setup of linear analysis}    & \multicolumn{4}{c}{$f-$mode frequency relative to the simulation } \\
\cline{2-9}
  & coordinates & boundary & lapse & conformal factor & Cowling & $\delta \alpha\neq 0 $ & $\delta \alpha,\delta\psi\neq 0  $ & full \\
\hline
M18m10$^{\rm a}$     & isotropic  & PNS, $\Delta P=0$    & $\exp\left(\Phi\right)$ & $\psi=1$     & low  & just & -    & -   \\
T19m20$^{\rm b}$     & isotropic  & shock, $\eta_r=0$    & $1+\Phi$  & $\psi=1$    & high & high & just & -   \\
m2.9INW$^{\rm c}$    & isotropic  & PNS, $\Delta P=0$    & $\exp\left(\Phi\right)$ & $\psi=1$     & high & just & low  & -   \\
m2.9IGR$^{\rm d}$    & isotropic  & PNS, $\Delta P=0$    & $\exp\left(\Phi\right)$ & $\psi\neq1$  & low  & low  & low  & -   \\
m2.9SGR$^{\rm e}$    & spherical  & PNS, $\Delta P=0$    & $\exp\left(\Phi\right)$ & $\exp(\Lambda)\neq 1$ & low  & -    & -    & low \\
         \hline\hline
\multicolumn{9}{l}{$^{\rm a}$  Fig.~5 in Ref.~\cite{MRBV2018}, 10\,$M_\odot$ progenitor with SFHo EOS.}\\
\multicolumn{9}{l}{$^{\rm b}$  Left panel of Fig.~12 in Ref.~\cite{TCPOF2019a}, 20\,$M_\odot$ progenitor with SFHo EOS.Their $^2g_1$-mode is called $f$-mode in this study.}\\
\multicolumn{9}{l}{$^{\rm c}$  Left panel of Fig.~\ref{fig:2D-spe}, 2.9\,$M_\odot$ He star with LS EOS. }\\
\multicolumn{9}{l}{$^{\rm d}$  Right panel of Fig.~\ref{fig:2D-spe}, 2.9\,$M_\odot$ He star with LS EOS.}\\
\multicolumn{9}{l}{$^{\rm e}$  Fig.~\ref{fig:ft}, 2.9\,$M_\odot$ He star with LS EOS.}\\
\end{tabular}
\label{tab:comp_sim}
\end{table}
%%%%%%%%%%%%%%%%%%%%%%

%\bibliographystyle{h-physrev} % for PrD
%\bibliography{mybib}

\begin{thebibliography}{999}
%%%%%%%%%%%%%%%%%%%%%%%%%%%%%%%%%%%%%%%%%%%%%%%%

%\bibitem{GW1}  % GW150914
%   B. P. Abbott et al. (LIGO Scientific Collaboration and Virgo Collaboration), Phys. Rev. Lett. {\bf 116}, 061102 (2016).

%\bibitem{GW2}  % GW151226
%   B. P. Abbott et al. (LIGO Scientific Collaboration and Virgo Collaboration), Phys. Rev. Lett. {\bf 116}, 241103 (2016).

%\bibitem{GW3}  % GW170104
%   B. P. Abbott et al. (LIGO Scientific Collaboration and Virgo Collaboration), Phys. Rev. Lett. {\bf 118}, 221101 (2017).

%\bibitem{GW4}  % GW170608
%   B. P. Abbott et al. (LIGO Scientific Collaboration and Virgo Collaboration), Astrophys. J. {\bf 851}, L35 (2017).

%\bibitem{GW5}  % GW170814
%   B. P. Abbott et al. (LIGO Scientific Collaboration and Virgo Collaboration), Phys. Rev. Lett. {\bf 119}, 141101 (2017).

\bibitem{GW6}  % GW170817
   B. P. Abbott et al. (LIGO Scientific and Virgo Collaboration), Phys. Rev. Lett. {\bf 119}, 161101 (2017).

\bibitem{EM}  % EM in GW170817
   B. P. Abbott et al. (LIGO Scientific and Virgo Collaboration), Astrophys. J. {\bf 848}, L12 (2017).

\bibitem{aso13} 
   Y. Aso, Y. Michimura, K. Somiya, M. Ando, O. Miyakawa, T. Sekiguchi, D. Tatsumi, and H. Yamamoto, \prd {\bf 88}, 043007 (2013).

%\bibitem{advv} S. Hild, A. Freise, M,  Mantovani, M., et al.\ 
%Classical and Quantum Gravity, {\bf 26}, 025005 (2009).

%\bibitem{KAGRA}
%   http://gwcenter.icrr.u-tokyo.ac.jp/en/researcher/parameter

\bibitem{punturo} 
   M. Punturo, H. L\"{u}ck, and M. Beker, in {\it A Third Generation Gravitational Wave Observatory: The Einstein Telescope},
   edited by M. Bassan, Advanced Interferometers and the Search for Gravitational Waves. Astrophysics and Space
   Science Library Vol. 404 (Springer, Cham, 2014).
   
\bibitem{CE}
   B. P. Abbott {\it et al}. (LIGO Scientific and Virgo Collaboration), Class. Quantum Grav. {\bf 34}, 044001 (2017).

%%%%%%%%%%%%%%
%\bibitem{Hayama15}
%   K. Hayama, T. Kuroda, K. Kotake, and T. Takiwaki, Phys.\ Rev.\ D {\bf 92}, 122001 (2015).

%\bibitem{Andresen2018}
%  H. Andresen, E. M\"{u}ller, H.-T. Janka, A. Summa, K. Gill, and M. Zanolin, Mon. Not. R. Astron. Soc. {\bf 486}, 2238 (2019).

%%%%%%%%%%%%%%
\bibitem{Murphy09}
    J.~W. Murphy, C.~D. Ott, and A. Burrows, Astrophys. J. {\bf 707}, 1173 (2009).

\bibitem{MJM2013}
   B. M\"{u}ller, H. -T. Janka, and A. Marek, Astrophys. J. {\bf 766}, 43 (2013).

\bibitem{Ott13} 
    C.~D. Ott, E. Abdikamalov, P. M{\"o}sta, R. Haas, S. Drasco, E. P. O'Connor, C. Reisswig, C. A. Meakin, and E. Schnetter,  Astrophys. J. {\bf 768}, 115 (2013).
   
\bibitem{CDAF2013}
   P. Cerd\'{a}-Dur\'{a}n, N. DeBrye, M. A. Aloy, J. A. Font, and M. Obergaulinger, Astrophys. J. Lett. {\bf 779}, L18 (2013).

\bibitem{Yakunin15}
    K.~N. Yakunin, A. Mezzacappa, P. Marronetti, S. Yoshida, S.~W. Bruenn, 
    W.~R. Hix, E.~J. Lentz, O.~E. Bronson Messer, J.~A. Harris, E. Endeve,
    J.~M. Blondin, and E.~J. Lingerfelt, Phys.\ Rev.\ D {\bf 92}, 084040 (2015).

\bibitem{KKT2016}
   T. Kuroda, K. Kotake, and T. Takiwaki, Astrophys. J. Lett. {\bf 829}, L14 (2016).

\bibitem{Andresen16} 
    H. Andresen, B. M\"{u}ller, E. M\"{u}ller, and H.-T. Janka, Mon. Not. R. Astron. Soc.  {\bf 468}, 2032, (2017).

\bibitem{Richers2017}
  S.~Richers, C.~D. Ott, E.~Abdikamalov, E.~O'Connor, and C.~Sullivan, Phys. Rev. D {\bf 95}, 063019 (2017).

\bibitem{Takiwaki2017}
  T.~Takiwaki and K.~Kotake, Mon. Not. R. Astron. Soc. {\bf 475}, L91 (2018).

\bibitem{OC2018}
   E. P. O'Connor and S. M. Couch, Astrophys. J. {\bf 865}, 81 (2018).

\bibitem{RMBVN19}
   D. Radice, V. Morozova, A. Burrows, D. Vartanyan, and H. Nagakura, Astrophys. J. Lett. {\bf 876}, L9 (2019).

\bibitem{VBR2019}
   D. Vartanyan, A. Burrows, and D. Radice, Mon. Not. R. Astron. Soc. {\bf 489}, 2227 (2019).

\bibitem{PM2020}
  J. Powell and B. M\"{u}ller, Mon. Not. R. Astron. Soc. {\bf 494}, 4665 (2020).

%%%%%%%%%%%%%
\bibitem{MRBV2018}
  V. Morozova, D. Radice, A. Burrows, and D. Vartanyan, Astrophys. J. {\bf 861}, 10 (2018).

\bibitem{SKTK2019}
   H. Sotani, T. Kuroda, T. Takiwaki, and K. Kotake, Phys. Rev. D {\bf 99}, 123024 (2019).

\bibitem{ST2020b}
   H. Sotani and T. Takiwaki, preprint arXiv:2008.00419.

\bibitem{TCPF2018}
   A. Torres-Forn\'{e}, P. Cerd\'{a}-Dur\'{a}n, A. Passamonti, and J. A. Font, Mon. Not. R. Astron. Soc. {\bf 474}, 5272 (2018).

\bibitem{TCPOF2019a}
   A. Torres-Forn\'{e}, P. Cerd\'{a}-Dur\'{a}n, A. Passamonti, M. Obergaulinger, and J. A. Font, Mon. Not. R. Astron. Soc. {\bf 482}, 3967 (2019).

\bibitem{TCPOF2019b}
   A. Torres-Forn\'{e}, P. Cerd\'{a}-Dur\'{a}n, M. Obergaulinger, B. M\"{u}ller and J. A. Font, Phys. Rev. Lett. {\bf 123}, 051102 (2019).

%%%%%%%%%%%%%

\bibitem{Mezzacappa20}
   A. Mezzacappa, P. Marronetti, R. E. Landfield, E. J. Lentz, K. N. Yakunin, S. W. Bruenn, W. R. Hix, O. E. B. Messer, E. Endeve, J. M. Blondin, and J. A. Harris, Phys. Rev. D {\bf 102}, 023027 (2020).

%%%%%%%%%%%%%

\bibitem{KS1999}
   K. D. Kokkotas and B. G. Schmidt, Living Rev. Relativ. {\bf 2}, 2 1999.

\bibitem{GNHL2011}
   M. Gearheart, W. G. Newton, J. Hooker, and B. -A. Li, Mon. Not. R. Astron. Soc. {\bf 418}, 2343 (2011).
   
\bibitem{SNIO2012}
   H. Sotani, K. Nakazato, K. Iida, and K. Oyamatsu, Phys. Rev. Lett. {\bf 108}, 201101 (2012);
   Mon. Not. R. Astron. Soc. {\bf 428}, L21 (2013);
   Mon. Not. R. Astron. Soc. {\bf 434}, 2060 (2013).

\bibitem{SIO2016}
   H. Sotani, K. Iida, and K. Oyamatsu, New Astron. {\bf 43}, 80 (2016);
   Mon. Not. R. Astron. Soc. {\bf 464}, 3101 (2017); {\bf 479}, 4735 (2018); 
   {\bf 489}, 3022 (2019).

\bibitem{AK1996}
   N. Andersson and K. D. Kokkotas, Phys.\ Rev.\ Lett.\ {\bf 77}, 4134 (1996).

\bibitem{AK1998}
   N. Andersson and K. D. Kokkotas, Mon.\ Not.\ R. Astron.\ Soc.\ {\bf 299}, 1059 (1998).

\bibitem{STM2001}
   H. Sotani, K. Tominaga, and K. I. Maeda, Phys.\ Rev.\ D {\bf 65}, 024010 (2001).

\bibitem{SH2003}
   H. Sotani and T. Harada, Phys.\ Rev.\ D {\bf 68}, 024019 (2003);
   H. Sotani, K. Kohri, and T. Harada, {\it ibid}.\ {\bf 69}, 084008 (2004).

\bibitem{SYMT2011}
   H. Sotani, N. Yasutake, T. Maruyama, and T. Tatsumi, Phys.\ Rev.\ D {\bf 83} 024014 (2011).

\bibitem{PA2012}
   A. Passamonti and N. Andersson, Mon.\ Not.\ R. Astron.\ Soc.\ {\bf 419}, 638 (2012).

\bibitem{DGKK2013}
   D. D. Doneva, E. Gaertig, K. D. Kokkotas, and C. Kr\"{u}ger, Phys.\ Rev.\ D {\bf 88}, 044052 (2013).

\bibitem{Sotani2020}
   H. Sotani, arXiv:2008.09839.
%%%%%%%%%%%

\bibitem{FMP2003}
   V. Ferrari, G. Miniutti, and J. A. Pons, Mon. Not. R. Astron. Soc. {\bf 342}, 629 (2003).

\bibitem{Burgio2011}
   G. F. Burgio, V. Ferrari, L. Gualtieri, and H.-J. Schulze,  Phys.\ Rev.\ D {\bf 84}, 044017 (2011).

\bibitem{FKAO2015}
   J. Fuller, H. Klion, E. Abdikamalov, and C. D. Ott, Mon.\ Not.\ R. Astron.\ Soc.\ {\bf 450}, 414 (2015).

\bibitem{ST2016}
   H. Sotani and T. Takiwaki, Phys.\ Rev.\ D {\bf 94}, 044043 (2016).

\bibitem{Camelio17}
   G. Camelio, A. Lovato, L. Gualtieri, O. Benhar, J. A. Pons, and V. Ferrari, Phys. Rev. D {\bf 96}, 043015 (2017).

\bibitem{SKTK2017}
   H. Sotani, T. Kuroda, T. Takiwaki, and K. Kotake, Phys.\ Rev.\ D {\bf 96}, 063005 (2017).

\bibitem{SS2019}
   H. Sotani and K. Sumiyoshi, Phys.\ Rev.\ D {\bf 100}, 083008 (2019).

\bibitem{WS2019}
   J. R. Westernacher-Schneider, E. O'Connor, E. O'Sullivan, I. Tamborra, M.-R. Wu, S. M. Couch, and F. Malmenbeck, 
   Phys. Rev. D {\bf 100}, 123009 (2019).

\bibitem{ST2020a}
   H. Sotani and T. Takiwaki, Phys.\ Rev.\ D {\bf 102}, 023028 (2020).
   
%%%%%%%%

\bibitem{YK1997}
   S. Yoshida and Y. Kojima, Mon.\ Not.\ R. Astron.\ Soc.\ {\bf 289}, 117 (1997).

%%%%%%%%

\bibitem{He29}
  T. J. Moriya, P. A. Mazzali, N. Tominaga, S. Hachinger, S. I. Blinnikov, T. M. Tauris, K. Takahashi, M. Tanaka, N. Langer, and P. Podsiadlowski, Mon. Not. R. Astron. Soc. {\bf 466}, 2085 (2017).

\bibitem{LS220}
  J. M. Lattimer and F. D. Swesty, Nucl. Phys. {\bf A535}, 331 (1991).

\bibitem{liebendoerfer2009}
M.~Liebendoerfer, S.~Whitehouse, and T.~Fischer, Astrophys. J. {\bf 698}, 1174 (2009).

\bibitem{takiwaki2014}
T.~Takiwaki, K.~Kotake, and Y.~Suwa, Astrophys. J. {\bf 786}, 83 (2014).

%%%%%%%%
%   3DnSNe
%%%%%%%%
  \bibitem{takiwaki2016}
  T.~Takiwaki, K.~Kotake, and Y.~Suwa, Mon. Not. R. Astron. Soc. {\bf 461}, L112 (2016).

\bibitem{oconnor2018}
  E.~O'Connor {\it et~al.}, J. Phys. G {\bf 45}, 104001 (2018).

\bibitem{kotake2018}
   K.~Kotake, T.~Takiwaki, T.~Fischer, K.~Nakamura, and G.~Mart\'{i}nez-Pinedo, Astrophys. J. {\bf 853}, 170 (2018).

\bibitem{nakamura2019}
   K.~Nakamura, T.~Takiwaki, and K.~Kotake, Publ. Astron. Soc. Jpn. {\bf 71}, 98 (2019).

\bibitem{sasaki2019}
   H.~Sasaki, T.~Takiwaki, S.~Kawagoe, S.~Horiuchi, and K.~Ishidoshiro, Phys. Rev. D {\bf 101}, 063027 (2020).

\bibitem{zaizen2019}
   M. Zaizen, J. F. Cherry, T. Takiwaki, S. Horiuchi, K. Kotake, H. Umeda, T. Yoshida, J. Cosmol. Astropart. Phys. {\bf 2020}, 111 (2020).

%%%%%%%%

\bibitem{note}
In this study, we adopt the numerical simulation data with non-rotating progenitor model. So, we consider that even one-dimensional (spherically symmetric) PNS models by averaging in the angular direction are sufficient ass the background models for the linear perturbations. But, if one would consider the situation by adopting the numerical simulation with the rotating progenitor and the rotational effect would become crucial in the PNS evolution, one must make a linear analysis on the two-dimensional background. 


%%%%%%%%

\bibitem{Marek06}
   A. Marek, H. Dimmelmeier, H.-T. Janka, E. M\"{u}ller, and R. Buras, Astron. Astrophys. {\bf 445}, 273 (2006).

%%%%%%%%


%\bibitem{horowitz2017}
%   C. J.~Horowitz, O. L.~Caballero, Z.~Lin, E.~O'Connor, and A.~Schwenk, Phys. Rev. C {\bf 95}, 025801 (2017).

%\bibitem{Demorest2010}
%   P. Demorest, T. Pennucci, S. Ransom, M. Roberts and J. Hessels, Nature {\bf 467}, 1081 (2010).

%\bibitem{Antoniadis2013}
%   J. Antoniadis {\it et al.}, Science {\bf 340}, 1233232 (2013).

%\bibitem{Annala18} 
%   E. Annala, T. Gorda, A. Kurkela, and A. Vuorinen, Phys. Rev. Lett. {\bf 120}, 172703 (2018).

%%%%%%%%%
%   progenitor
%%%%%%%%%
%\bibitem{WH07}  % WH07
%   S. E. Woosley and A. Heger, Phys. Rep. {\bf 442}, 269 (2007).

%\bibitem{bruenn2013}
%   S.~W. Bruenn {\it et~al.}, Astrophys. J. {\bf 767}, L6 (2013).

%\bibitem{dolence2015}
%   J.~C. Dolence, A.~Burrows, and W.~Zhang, Astrophys. J. {\bf 800}, 10 (2015).

%\bibitem{summa2016}
%   A.~Summa, F. Hanke, H.-T. Janka, T. Melson, A. Marek, and B. M\"{u}ller,  Astrophys. J. {\bf 825}, 6 (2016).

%\bibitem{suwa2016}
%   Y.~Suwa, S.~Yamada, T.~Takiwaki, and K.~Kotake, Astrophys. J. {\bf 816}, 43 (2016).

%\bibitem{pan2016}
%   K.-C. Pan, M.~Liebend\"{o}rfer, M.~Hempel, and F.-K. Thielemann, Astrophys. J. {\bf 817}, 72 (2016).

%\bibitem{skinner2016}
%   M.~A. {Skinner}, A.~{Burrows}, and J.~C. {Dolence}, Astrophys. J. {\bf 831}, 81 (2016).

%\bibitem{bruenn2016}
%   S.~W. Bruenn {\it et~al.}, Astrophys. J. {\bf 818}, 123 (2016).

%\bibitem{oconnor2018b}
%   E.~P. O'Connor and S.~M. Couch, Astrophys. J. {\bf 854}, 63 (2018).

%\bibitem{just2018}
%   O.~Just, R. Bollig, H.-T. Janka, M. Obergaulinger, R. Glas, S. Nagataki, Mon. Not. R. Astron. Soc. {\bf 481}, 4786 (2018).

%\bibitem{glas2019}
%   R.~Glas, O.~Just, H.~T. Janka, and M.~Obergaulinger, Astrophys. J. {\bf 873}, 45 (2019).

%\bibitem{melson2015}
%   T.~Melson, H.-T. Janka, R. Bollig, F. Hanke, A. Marek, and B. M\"{u}ller, Astrophys. J. {\bf 808}, L42 (2015).

%\bibitem{bollig2017}
%   R.~Bollig, H.-T. Janka, A. Lohs, G. Mart\'{i}nez-Pinedo, C. J. Horowitz, and T. Melson, Phys. Rev. Lett. {\bf 119}, 242702 (2017).

%\bibitem{vartanyan2018}
%   D.~Vartanyan, A.~Burrows, D.~Radice, M.~A. Skinner, and J.~Dolence, Mon. Not. R. Astron. Soc. {\bf 477}, 3091 (2018).

%%%%%%%%

\bibitem{LD83}
   L. Lindblom and S. Detweiler, Astrophys. J. Suppl. Ser. {\bf 53}, 73 (1983); 
   S, Detweiler and L. Lindblom, Astrophys. J. {\bf 292}, 12 (1985).

\bibitem{LNS93}
   M. Leins, H.-P. Nollert, and M. H. Soffel, Phys. Rev. D {\bf 48}, 3467 (1993).

\bibitem{Chandrasekhar83}
   S. Chandrasekhar, {\it The Mathematical Theory of Black Holes} (Clarendon, Oxford, 1983), Chap. 4.
   
   
   


\end{thebibliography}
%%%%%%%%%%%%%%%%%%%%%%%%%%%%%%%%%%%%%%%%%%%%%%%%

\end{document}